# Double bit in-plane magnetic skyrmions on a track

Kyoung-Woong Moon†, Jungbum Yoon†, Changsoo Kim, and Chanyong Hwang*

Spin Convergence Research Team, Korea Research Institute of Standards and Science,

Daejeon 34113, Republic of Korea

*Correspondence to C.H. (cyhwang@kriss.re.kr)

†These authors equally contribute to this work.

A magnetic skyrmion usually refers to a twisted spin texture surrounded by uniformly aligned out-of-plane spins[1–6]. The invariance of the topological charge under any continuous transformations of the spin textures[1,2,7–9] leads to the robustness of a magnetic skyrmion against external perturbations, making it ideal to use skyrmions as information carriers[1,2,5,6]. To date, most magnetic skyrmion studies have been focused on perpendicularly magnetized systems, where the skyrmion topological number is determined by the relative orientation between the core and outer perpendicular magnetization directions. Here we show that there also exists a new type of magnetic skyrmion with surrounding spins to be uniformly aligned to the in-plane direction. Different from the conventional perpendicular magnetic skyrmions[1,2,6,10], the in-plane magnetic skyrmions with opposite signs of topological charges can inherently coexist[11].

**Moreover, the in-plane skyrmions of opposite charge move together by an electric current and exhibit opposite spin Hall effect. These findings demonstrate the inherent possibility of a double-bit transfer in a single magnetic wire that is not possible in a perpendicularly magnetized system.**

It is a common assumption that two-dimensional magnetic skyrmions exist in system with perpendicular magnetic anisotropy (PMA)[1–6] which prefers only up or down magnetization directions. A skyrmion can be regarded as a topological defect in a uniform perpendicularly magnetized state. Figure 1a shows a typical spin state of a skyrmion in uniform magnetization along the $-z$ (down) direction (black arrows). At the centre of the skyrmion, there is a core magnetization along the $+z$ (up) magnetization (white arrows) opposite to the outer background magnetization. Between the core and the outer region, there is a domain wall where the magnetization has to rotate in space to connect the up and down magnetization. As a result, pure in-plane magnetization is required at the centre of the magnetization rotation (rainbow colour arrows) that wraps the $+z$ core.

To form the skyrmion state, the warping magnetization should have a rotational symmetry. A sufficient effective magnetic field for holding the magnetization can be generated by the Dzyaloshinskii–Moriya interaction (DMI)[12,13], which generates a DMI field ($\mathbf{H}_{DMI}$) where the magnetization gradient occurs. In this paper, we assume the interfacial-type DMI (iDMI)[2,11,14,15]; thus, $\mathbf{H}_{DMI}$ prefers a certain chiral spin texture such as $(-z)$-$(+x)$-$(+z)$-$(-x)$-$(-z)$ along the $x$-axis. If we consider the same chiral spin configuration to the $y$-axis also, we come up the basic skyrmion texture (Fig. 1a). Figure 1b presents the other skyrmion state with the outer magnetization $+z$. In this situation, the core magnetization becomes $-z$ and the warping in-plane magnetization should be reversed to maintain the same chiral spin configuration because of iDMI. Regardless of the core direction, $\mathbf{H}_{DMI}$ causes the skyrmion to be symmetrically centred at the skyrmion core. Thus, the skyrmion in PMA looks like a circular bubble domain.

Without considering the spin texture at the boundary, this circular domain structure has been claimed to be the skyrmion. However, the exact meaning of skyrmion originates from the topology[7,8,16,17]. The topology does not determine the detailed configuration; i.e., the size and the shape of an arbitrary magnetic texture cannot be the criterion for claiming the skyrmion. The only criterion is the topological charge, defined as $Q = (1/4\pi) \int \mathbf{m} \cdot [\partial_x \mathbf{m} \times \partial_y \mathbf{m}] \, dxdy$. Here, $Q$ is the skyrmion charge and $\mathbf{m}$ ($=(m_x, m_y, m_z)$) is the normalized magnetization vector at a certain position $(x, y)$. When $Q$ is quantized at the value $\pm 1$, we can refer to this state as a skyrmion (for example, $Q = +1$ for Fig. 1a and $Q = -1$ for Fig. 1b).

The skyrmion charge is conserved under continuous transformations such as translation and rotation[11]. Translation is directly related to skyrmion motion[1,2]; we already know that a position change of a skyrmion does not change the skyrmion charge. This gives us a clue for building the skyrmion in an in-plane magnetic anisotropy system. Now, let us consider a vector rotation by an angle of 90° about the *y*-axis. The inset of Fig.1a shows the result of rotation of the PMA skyrmion ($Q = +1$). The outer magnetization changes from $-z$ to $+x$ and the core magnetization changes from $+z$ to $-x$. After the rotation, the $\pm z$ magnetizations appear at the side of the skyrmion. Similar vector rotation is possible for the opposite skyrmion ($Q = -1$). In this case, the rotation direction should be reversed and the rotation makes the outer magnetization to be $+x$. Note that a proper magnetization rotation can cause the same outer magnetization, regardless of the skyrmion charges.

Figure 1c shows skyrmions with opposite $Q$ signs in an in-plane magnetic anisotropy(IMA) system. Because of the same outer magnetization, two opposite skyrmions can coexist with the same background outer magnetization. To obtain this magnetization state, we set the inset of Fig. 1a and 1b as initial states and relax the spin configuration to find a local energy minimum state by a micromagnetic simulation programme[18]. We can see that the in-plane skyrmion has a spin structure deformed from the initial states. The spin configuration has a mirror symmetry about the *x*-axis but no symmetry about the *y*-axis. The overall shape of the in-plane skyrmion is a vortex and an anti-vortex pair. Magnetization rotation and the local deformation of the spin configuration might change the skyrmion charge. With simple vortex descriptions we can verify the in-plane skyrmion charge[19–22]. In magnetic vortex studies, the topological charge of a vortex is $pn/2$, where $p$ is the polarity of the vortex core; $p$ is $+1(-1)$ with a $+z(-z)$ core; and $n$ represents the winding number of the vortex, where $n$ is $+1(-1)$ with vortex (anti-vortex). The IMA skyrmion consists of the vortex and anti-vortex pair and the polarities of the two cores are opposite. As a result, the sum of topological charges of an IMA skyrmion is ±1 (Fig. 1c).

Now, we will show the detailed shape of skyrmions in an IMA system. In a PMA system, the skyrmion looks like a circular domain. The relaxed magnetization state of a PMA skyrmion with a simplified colour code is shown in Fig. 2a. The different colours show the region where the major magnetization direction is one of $\pm x$, $\pm y$, and $\pm z$. In this PMA skyrmion, the position where $m_z = 0$ is considered to be an edge defining the skyrmion shape (the cyan line in Fig. 2a). By the same description, Fig. 2b shows the structure of the relaxed in-plane skyrmion (see Methods for detailed conditions). In this case, the position with $m_x = 0$ is an edge of the in-plane skyrmion. The insets of Fig. 2b represent the magnetization configurations on a grey colour scale; white ($m_i = +1$) and black ($m_i = -1$). The vortex with the $+z$ core is almost symmetric, but the anti-vortex with the $-z$ core is compressed along the $x$-direction.

This IMA skyrmion is mainly determined by the competition between iDMI and the demagnetization field. To form the skyrmion texture in IMA, it requires perpendicular magnetizations, but such a perpendicular magnetization produces a magnetic charge on the surface of the magnetic material that generates an opposite magnetic field inside the magnet. This is the demagnetization field ($\mathbf{H}_{\text{demag}}$) and its strength is equal to the saturation magnetization ($M_S$). In an IMA system, the change of the in-plane magnetization in space can produce an effective perpendicular magnetic field because of the iDMI. A simplified magnetization state is shown in the lower inset of Fig. 2c. When the magnetization is uniform along the $x$-axis and changes from $+y$ to $-y$ in a length scale ($l_S$) along the $y$-direction, the strength of the iDMI field ($H_{\text{DMI}}$) at the centre is $\frac{\pi D_{\text{ind}}}{2\mu_0 M_S l_S}$ (ref. 23), where $D_{\text{ind}}$ is the iDMI energy density and $\mu_0$ is the permeability. Therefore, if $|H_{\text{DMI}}|>M_S$, a perpendicular magnetization can be stabilized in IMA. We know that magnetization change needs the exchange length $\sqrt{2A/(\mu_0 M_S^2)}$, where $A$ is the exchange stiffness constant[22]. Inserting the exchange length as $l_S$ results in $M_S < 1.11|D_{\text{ind}}|/\sqrt{\mu_0 A}$ (see Methods for the iDMI field).

The phase diagram of magnetization states as functions of $D_{\text{ind}}$ and $M_S$ are shown in Fig. 2c. Each magnetization state was obtained by relaxing the initial magnetization state that contains a single IMA skyrmion in $+x$ outer magnetization such as in Fig. 2b. Thus, each image represents a local energy minimum state as well as the condition where the IMA skyrmion survives (white boxed images). The black line represents the position where $M_S = 1.11|D_{\text{ind}}|/\sqrt{\mu_0 A}$ (called the stripe condition). An increment in $D_{\text{ind}}$ or decrement in $M_S$ from this line corresponds to a $H_{\text{DMI}}$ value larger than $M_S$, which produces full stripe domains.

This fits well with our expectation because the condition for a stripe domain is derived from a one-dimensional spin structure (the lower inset of Fig. 2c).

An IMA skyrmion can exist near the stripe condition but the required $D_{\text{ind}}$ is slightly smaller than that for the stripe condition. The upper inset of Fig. 2c depicts a simplified magnetic configuration of the IMA skyrmion. It has $+z$ magnetization stabilized by the perpendicular iDMI field. This iDMI field has an additional contribution because of the magnetization variation from $+x$ to $-x$ along the $x$-direction. Thus, the iDMI field is double that of the stripe states, and therefore the required $D_{\text{ind}}$ should be small. At $-z$ magnetization of the IMA skyrmion, $m_y$ variations along the $y$-axis generate the iDMI field assisting $\mathbf{H}_{\text{demag}}$. These two fields make the $-z$ magnetization unstable, and thus these two fields must be compensated by the other iDMI field generated by the $m_x$ variation in the $x$-axis. We know that the iDMI field is inversely proportional to $l_S$, and thus the area of $-z$ magnetization should have been reduced along the $x$-direction. This is the main origin of the asymmetric shape of the IMA skyrmion.

Now we can apply an electric current to induce the motion of the IMA skyrmion. The electric current-induced skyrmion motion is an important technical issue in the application area[1,2,5,6]. There are two well-known mechanisms for the motions. One is induced by the current flowing in the magnet and we call this spin-magnetization-transfer torque (SMT)[2,24–27]. The other comes from the current in the attached heavy metal layers and is known as spin–orbit torque (SOT)[2,26,28]. Each effect inherently has a vector property; SMT has the vector $\mathbf{u}$ representing the current direction and strength and SOT has the vector $\boldsymbol{\sigma}$ representing a pumped spin direction. Generally, $\mathbf{u}$ and $\boldsymbol{\sigma}$ are on the $xy$-plane and can have arbitrary direction; thus, we check the skyrmion velocities as a function of angle difference between the outer magnetization ($\mathbf{m}_{\text{out}}$) and each of $\mathbf{u}$ and $\boldsymbol{\sigma}$ (see Methods). This is a property distinct from the PMA skyrmion because the $\mathbf{m}_{\text{out}}$ of the PMA skyrmion is the $\pm z$ magnetization that has no angular dependence on $\mathbf{u}$ and $\boldsymbol{\sigma}$.

Figure 3 shows the results of the IMA skyrmion velocity in infinite films. We set $\mathbf{m}_{\text{out}}$ as the $-y$ magnetization and change the angle of the SMT effect ($\varphi_{\mathbf{u}}$) and of the SOT effect ($\varphi_{\boldsymbol{\sigma}}$). The red and blue arrows in the lower panel depict skyrmion velocities in the $xy$-plane. The starting points of these arrows define $\varphi_{\mathbf{u}}$ and $\varphi_{\boldsymbol{\sigma}}$ with respect to $+y$. We can see a weak velocity dependence on $\varphi_{\mathbf{u}}$, but a strong dependence on $\varphi_{\boldsymbol{\sigma}}$. The non-adiabaticity of SMT ($\beta$) determines the transverse components of skyrmion velocity with respect to $\mathbf{u}$ (Fig. 3a). In SOT motion, a larger damping constant ($\alpha$) increases the transverse speed with respect to $\boldsymbol{\sigma} \times \hat{z}$. Such a transverse motion is known as the skyrmion Hall effect[10,17]. Both these velocity results and the skyrmion Hall effect seem to be quite complicated. However, Thiele's equations are good for predicting the skyrmion velocity (dashed lines and dots), which is useful for describing the motions of well-structured magnetic configurations[2,7,11,25,27,29]. The Thiele approach reveals that the asymmetric shape of the IMA skyrmion is the origin of the complicated skyrmion velocity (see Methods).

Finally, we show that skyrmions having opposite $Q$ values can move together. Figure 4a shows the skyrmion motion in a wire structured track by SMT and Fig. 4b shows the motion by SOT (see Supplementary Movie). Initially, two IMA skyrmions are placed in a uniform $-y$ magnetization. At the beginning of the motion, opposite $Q$ produces opposite skyrmion Hall effects, but when the skyrmions meet wire edges, the edges generate a repulsive force to the skyrmions[25]. As a result, two skyrmions with opposite $Q$ values can move together along the track direction. We know that the skyrmions act as a data bit, and thus this result means that we can transfer double data bits in a single wire track.

In summary, we show the possibility of skyrmion structures in the in-plane magnetization system using the topological property of the skyrmion, namely, charge conservation under continuous transformation. The in-plane skyrmion can be stabilized by DMI because this interaction can hold the perpendicular magnetization. The important property of this in-plane skyrmion is that skyrmions having opposite charge can coexist in the same outer background magnetization. This is a distinct feature of the in-plane skyrmions over the perpendicular system. This compatibility of opposite skyrmions enables double-bit transfer on a single track. Furthermore, an operator based on these two bits could be possible upon a logic gate (the readout head).

**Methods**

**Micromagnetic simulations.** We performed the micromagnetic simulations using the mumax code[18]. In Fig. 2, we set the following parameters: the exchange stiffness constant ($A$=1.3 × 10$^{-11}$ J m$^{-1}$), the interfacial–DMI energy density ($D_{\text{ind}}$=2×10$^{-3}$ J m$^{-2}$), the saturation magnetization ($M_S$=720×10$^3$ A m$^{-1}$), and the magnetic layer thickness ($t_M$=1 nm) (see Supplementary Fig. 1 for a phase diagram with different $t_M$). Only for Fig. 2a, we used the PMA energy density ($K_z$=600×10$^3$ J m$^{-3}$) to make the perpendicular system. The cell size was 0.5×0.5×1 nm$^3$ (Fig. 2a, b) and 2.5×2.5×1 nm$^3$ (Fig. 2c). Each image size in Fig. 2c corresponds to 500×500 nm$^2$. The periodic boundary condition (PBC) is used along the –x- and y-axes. We get each magnetization state by relaxation. In Figs 3 and 4, we performed time-dependent simulations with $A$ (1.3×10$^{-11}$ J m$^{-1}$), $D_{\text{ind}}$ (1.6×10$^{-3}$ J m$^{-2}$), $M_S$ (560×10$^3$ A m$^{-1}$), $t_M$ (1 nm). We used the damping constant $\alpha$ (0.3) unless it is specified separately. To hold the outer magnetization to the −y direction, a small IMA along the y-direction

($K_y$=5×10³ J m⁻³) is introduced in Fig. 3 and Fig. 4 (see Supplementary Fig. 2 for the SMT-induced motions with $K_y$=0). We set the strength of SMT (|**u**|) as 5 m s⁻¹ for Fig. 3a and 10 m s⁻¹ for Fig. 4a. We used the strength of SOT ($\tau_d$) as 4 mT for Fig. 3b and 8 mT for Fig. 4b. The cell size is 2.5×2.5×1 nm³ in Figs 3 and 4.

**DMI and its effective magnetic field.** A full description for the effective magnetic field for the interfacial-type DMI is $\mathbf{H}_{\text{DMI}} = -2D_{\text{ind}}/(\mu_0 M_S)[(\nabla \cdot \mathbf{m})\hat{z} - \nabla m_z]$ (ref. 2). This equation means that the $m_z$ variation along the $i$-axis produces an $i$-directional effective magnetic field. Here, $i$ is $x$ or $y$. In addition, the variation of $m_i$ along the $i$-axis generates a perpendicular iDMI field. Note that the iDMI field is proportional to the derivative of magnetization in position; the magnetization change in a shorter length scale generates a larger iDMI field. In the PMA system, the typical length scale is known as the domain wall width. The magnetization change from $\pm z$ to $\mp z$ magnetization requires a wall width ($\Delta_0$) and $\Delta_0 = \sqrt{A/\{K_z - (\mu_0/2)M_S^2\}}$ [14,15,23]. This length scale is determined by the energy cost of the exchange energy and the anisotropy energy. In the PMA system, the anisotropy energy comes from in-plane magnetization of the domain wall. If we convert this approach for our IMA system ($K_z = 0$), the perpendicular magnetization requires the demagnetization energy that changes the sign of the demagnetization energy part in $\Delta_0$. Therefore, $\Delta_0$ becomes the exchange length ($\sqrt{2A/(\mu_0 M_S^2)}$). In the PMA system, the iDMI field strength at the centre of the domain wall is simply $\pi D_{\text{ind}}/(2\mu_0 M_S \Delta_0)$ [23]. Note that, bulk-type DMI is also possible. We obtain the phase diagram of the bulk type of DMI in Supplementary Fig. 3.

**Thiele's equations.** The dynamics of local magnetization are described by the Landau–Lifshitz–Gilbert (LLG) equation as follows:

$$\dot{\mathbf{m}} = -\gamma_0 \mathbf{m} \times \mathbf{H}_{\text{eff}} + \alpha \mathbf{m} \times \dot{\mathbf{m}} - (\mathbf{u} \cdot \nabla)\mathbf{m} + \beta \mathbf{m} \times [(\mathbf{u} \cdot \nabla)\mathbf{m}]$$

$$+\gamma_0 \tau_d \mathbf{m} \times (\boldsymbol{\sigma} \times \mathbf{m}) + \gamma_0 \tau_f \mathbf{m} \times \boldsymbol{\sigma}. \tag{1}$$

Here, $\mathbf{m}$ is the local magnetization, $\dot{\mathbf{m}}$ is the time derivative of $\mathbf{m}$, $\gamma_0$ is the gyromagnetic ratio, $\mathbf{H}_{\text{eff}}$ is the effective magnetic field, $\alpha$ is the damping constant, $\mathbf{u}$ represents the SMT effect, $\beta$ is the non-adiabaticity of SMT, $\tau_d$ is the strength of damping-like SOT, $\boldsymbol{\sigma}$ is the pumped spin direction of SOT, and $\tau_f$ is the strength of field-like SOT. In this paper, we set $\mathbf{H}_{\text{eff}}$=0 and $\tau_f$=0 because we do not discuss the magnetic field and the field-like effects. Then, the LLG equation becomes,

$$\dot{\mathbf{m}} = \alpha \mathbf{m} \times \dot{\mathbf{m}} - (\mathbf{u} \cdot \nabla)\mathbf{m} + \beta \mathbf{m} \times [(\mathbf{u} \cdot \nabla)\mathbf{m}] + \gamma_0 \tau_d \mathbf{m} \times (\boldsymbol{\sigma} \times \mathbf{m}). \tag{2}$$

The Thiele approach is useful when the magnetization configuration is conserved during the motions. This means $\mathbf{m}(x, y) = \mathbf{m}_0(x - V_x t, y - V_y t)$, where $V_i$ is $i$-directional speed and $t$ is the time. To get Thiele's equation, we have to convert each torque term in the LLG equation to a speed term by calculating $\int (\text{LLG equation}) \cdot (\mathbf{m} \times \partial_i \mathbf{m}) dx dy$ [2,7,11,25,27,29]. Here, $i$ is $x$ or $y$. Using $\dot{\mathbf{m}} = -V_x \partial_x \mathbf{m}_0 - V_y \partial_y \mathbf{m}_0$ and vector calculus identities, the equations $\mathbf{A} \cdot (\mathbf{B} \times \mathbf{C}) = \mathbf{B} \cdot (\mathbf{C} \times \mathbf{A}) = \mathbf{C} \cdot (\mathbf{A} \times \mathbf{B})$ and $(\mathbf{A} \times \mathbf{B}) \cdot (\mathbf{C} \times \mathbf{D}) = (\mathbf{A} \cdot \mathbf{C})(\mathbf{B} \cdot \mathbf{D}) - (\mathbf{B} \cdot \mathbf{C})(\mathbf{A} \cdot \mathbf{D})$, produce two coupled equations. When only the SMT drives the skyrmion ($\tau_d$=0), Thiele's equations are,

$$-G(u_y - V_y) = D_{xx}(\beta u_x - \alpha V_x), \tag{3a}$$

$$G(u_x - V_x) = D_{yy}(\beta u_y - \alpha V_y). \tag{3b}$$

Here, $G = -4\pi Q = -\int \mathbf{m}_0 \cdot (\partial_x \mathbf{m}_0 \times \partial_y \mathbf{m}_0) dx dy$, $D_{ij} = \int \partial_i \mathbf{m}_0 \, \partial_j \mathbf{m}_0 \, dx dy$, and $\mathbf{u}=(u_x, u_y)$. Equation (3a) and (3b) describe the skyrmion velocities in Fig. 3a. If we consider that only the SOT drives the skyrmion ($\mathbf{u}$=0), Thiele's equations are,

$$GV_y = -\alpha D_{xx}V_x + w_x, \tag{4a}$$

$$-GV_x = -\alpha D_{yy}V_y + w_y. \tag{4b}$$

Here, $w_i$ ($=\gamma_0\tau_d \int (\boldsymbol{\sigma} \times \mathbf{m}_0) \cdot \partial_i \mathbf{m}_0 dxdy$) is the speed contribution because of the damping-like SOT. Equation (4) well describes the skyrmion motions shown in Fig. 3b. Note that it is $\partial_x\mathbf{m}_0 \neq \partial_y\mathbf{m}_0$ in the IMA skyrmion that makes the velocity complicated. Inserting $V_y=0$ in the above equations simply results in $V_x = (\beta/\alpha)u_x$ and $V_x = 1/(\alpha D_{xx})w_x$. These speed equations are useful to predict the skyrmion speeds in the confined wire structure as shown in Fig. 4.

**Acknowledgements**

This work was supported by the Future Materials Discovery Program through the National Research Foundation of Korea (No. 2015M3D1A1070467) and a National Research Council of Science & Technology (NST) grant (No. CAP-16-01-KIST) from the Korean government (MSIP).

**Author contributions:** K.W.M. designed the study. J.Y. and K.-W.M. performed micromagnetic simulations. C.H. directed and supported the project. J.Y., K.-W.M., and C.K. analysed the data. K.-W.M., and J.Y. drew the figures. K.-W.M., J.Y., C.K., and C.H. wrote the manuscript. All authors discussed the results and commented on the manuscript.

**Competing interests:** The authors declare no competing financial interests.

**Additional information** Supplementary information and Supplementary movie

**Figure captions**

**Figure 1| Skyrmion states in a perpendicular system and in an in-plane system. a,** Skyrmion with $Q = +1$. Arrows represent local magnetization; white and black colours show the *z*-directional magnetization; rainbow colours represent magnetization in the *xy*-plane. Inset shows a rotated magnetization state. **b,** Skyrmion with $Q = -1$. **c,** Coexistence of skyrmions in an in-plane magnetization system. The lower panel shows simplified magnetization states.

**Figure 2| Skyrmion structure and skyrmion phase diagram. a,** Simplified magnetization structure for a skyrmion in PMA. Colours show the direction ($\pm x, \pm y, \pm z$) of magnetization components that is dominant. The grey arrows depict the magnetization. **b,** Simplified magnetization structure for a skyrmion in IMA. Tree insets are images of $m_x$, $m_y$, and $m_z$ components with grey scale (white is +1 and black is −1). The scale bar is 20 nm. **c,** Relaxed magnetization states with respect to $D_{\text{ind}}$ and $M_S$. Insets are the simplified magnetic configurations of a skyrmion and a stripe. Purple arrows are the demagnetization field and cyan arrows are the iDMI field. Thick black line is the stripe condition (see Methods for simulation details).

**Figure 3| Skyrmion velocity and spin–torque angle in an infinite film. a,** SMT-driven skyrmion. **b,** SOT-driven skyrmion. We set the outer magnetization as $-y$. The upper images show the schematic magnetization state and angle definitions for **u** and **σ.** In the lower images, red and blue arrows represent the velocities of skyrmion motion in the *xy*-plane. The strengths of SMT and SOT are fixed and only the directions are changed. $\varphi_{\mathbf{u}}$ and $\varphi_{\boldsymbol{\sigma}}$ show the angles. Two black concentric circles depict |**u**| as the radius difference. Dashed lines and dots are the expected value from Thiele's approach (see Methods for simulation details).

**Figure 4| Skyrmion motions by spin torques in wires. a,** SMT-driven skyrmions. **b,** SOT-driven skyrmions. We set the outer magnetization is $-y$. Only $m_z$ components are shown with a grey scale (white is +1 and black is −1). **u** and **σ** are shown in the figure. Skyrmions are overlapping in a single image with a fixed time step (30 ns). The scale bar is 500 nm. White dashed lines show the trajectories of the motion. Red arrows are the moving direction (see Methods for simulation details and Supplementary Movie). The periodic boundary condition is used along the *x*-direction.

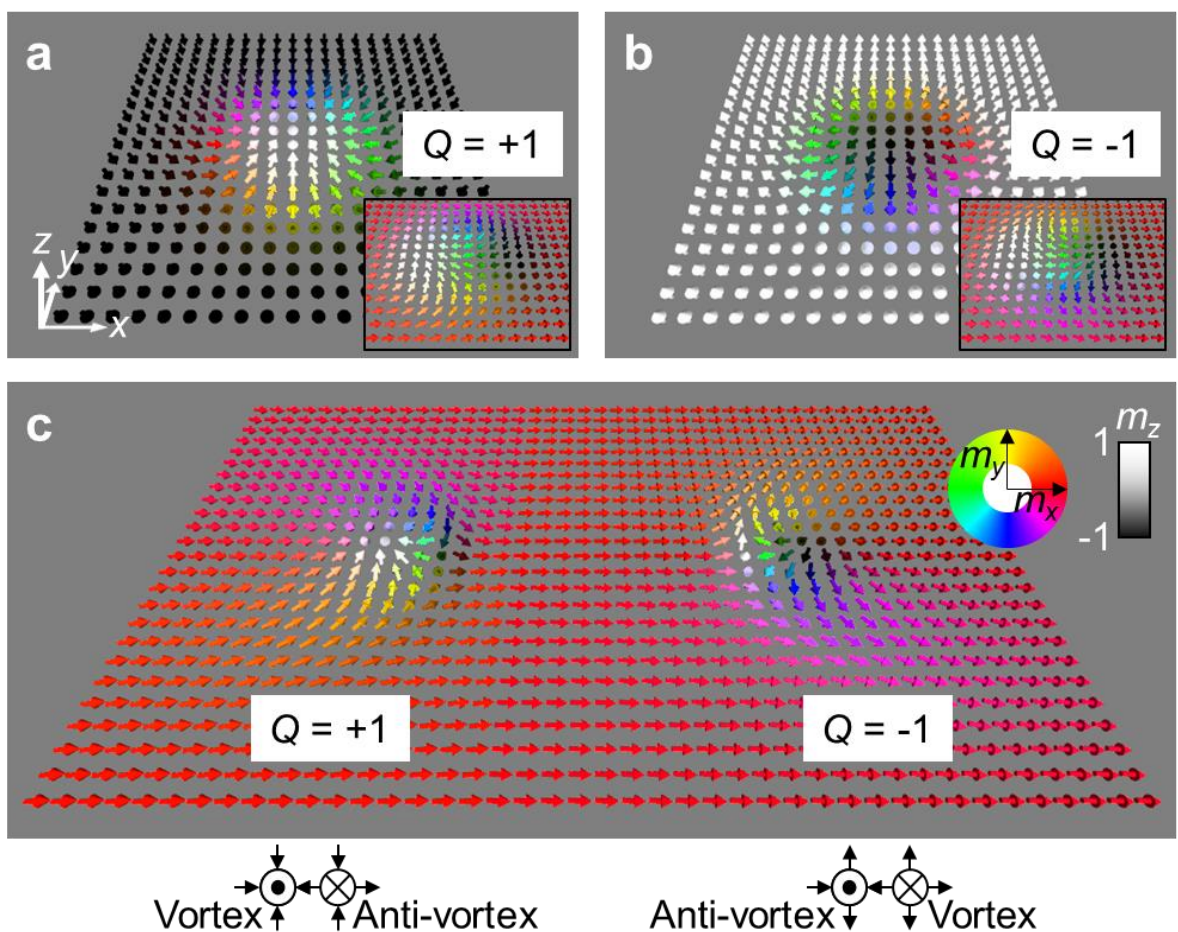

FIG. 1

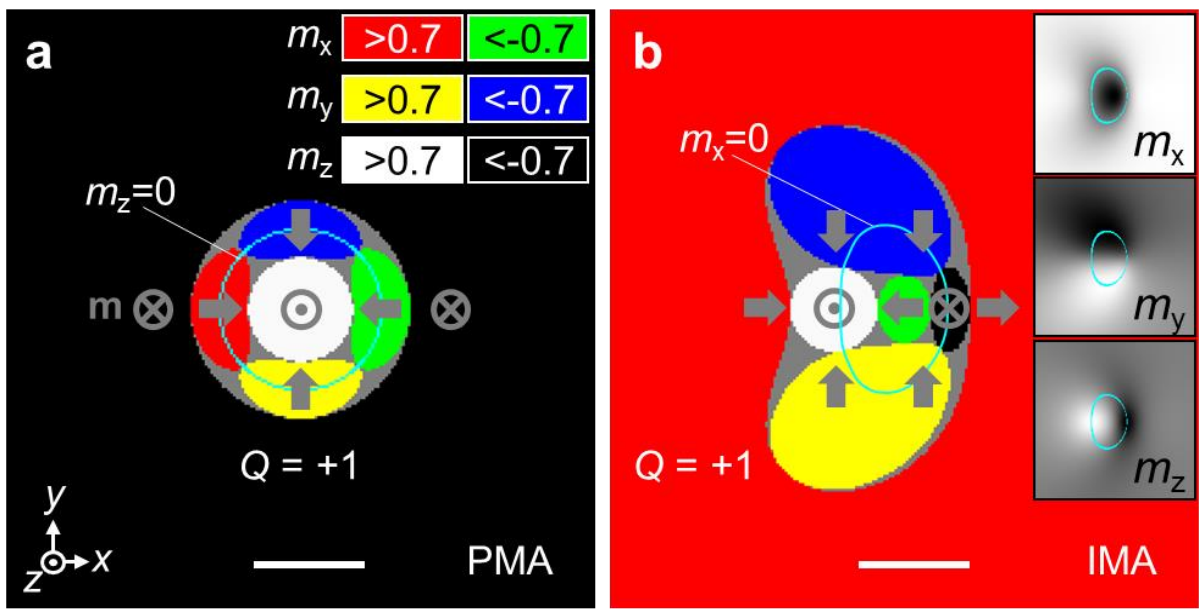

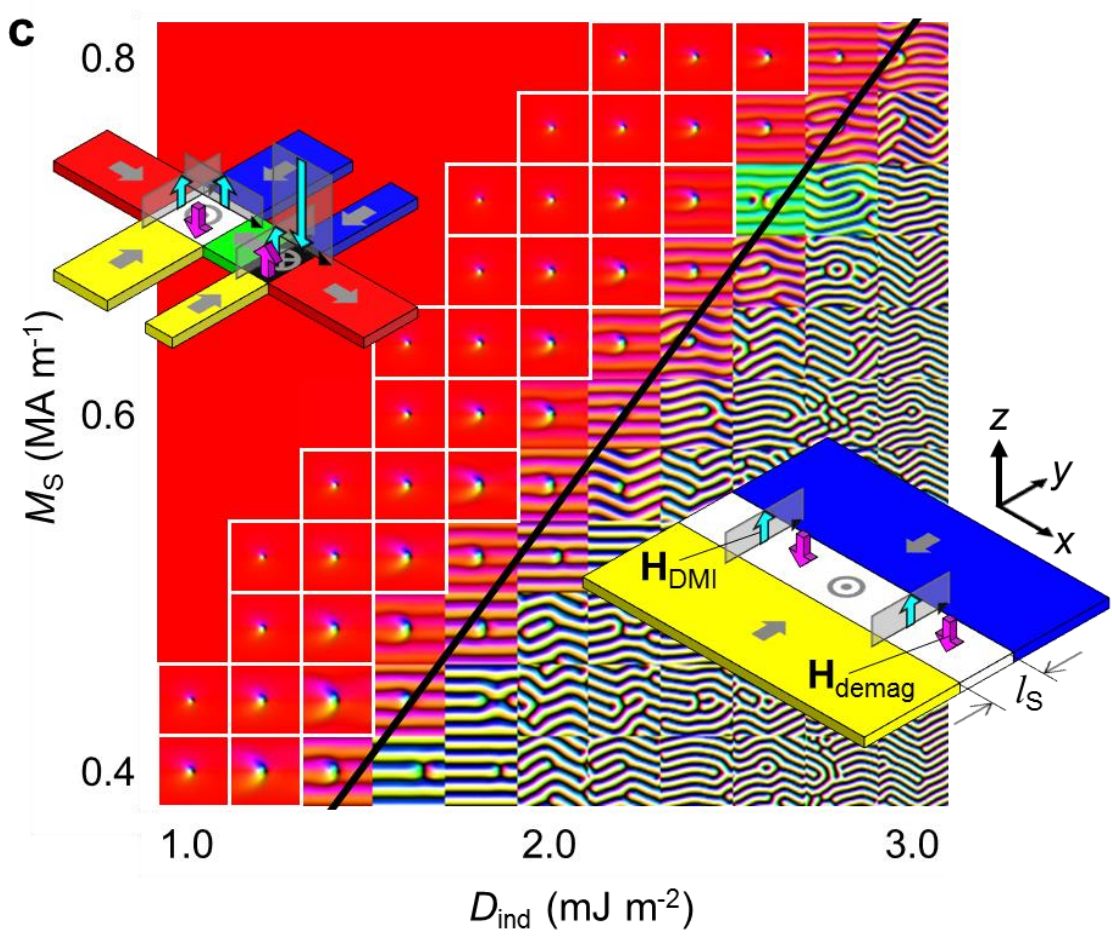

FIG. 2

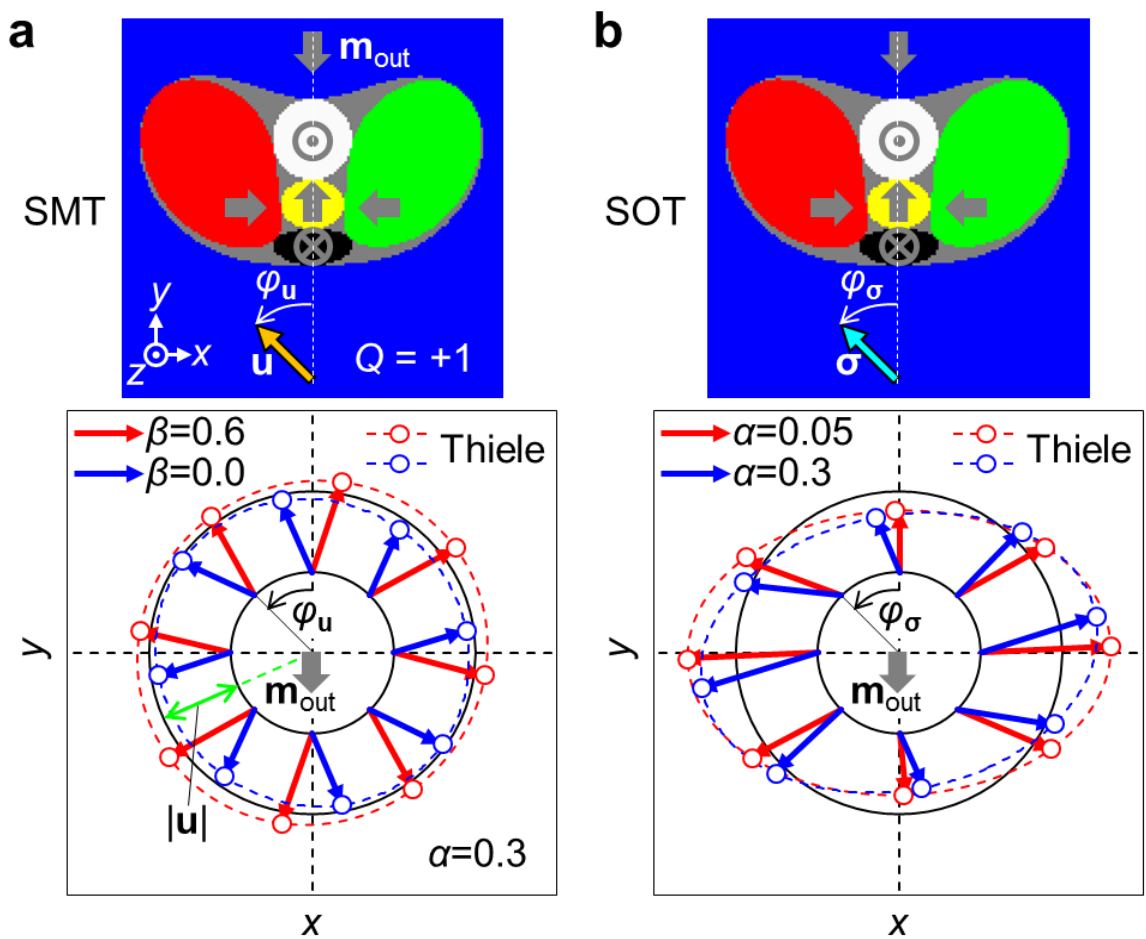

FIG. 3

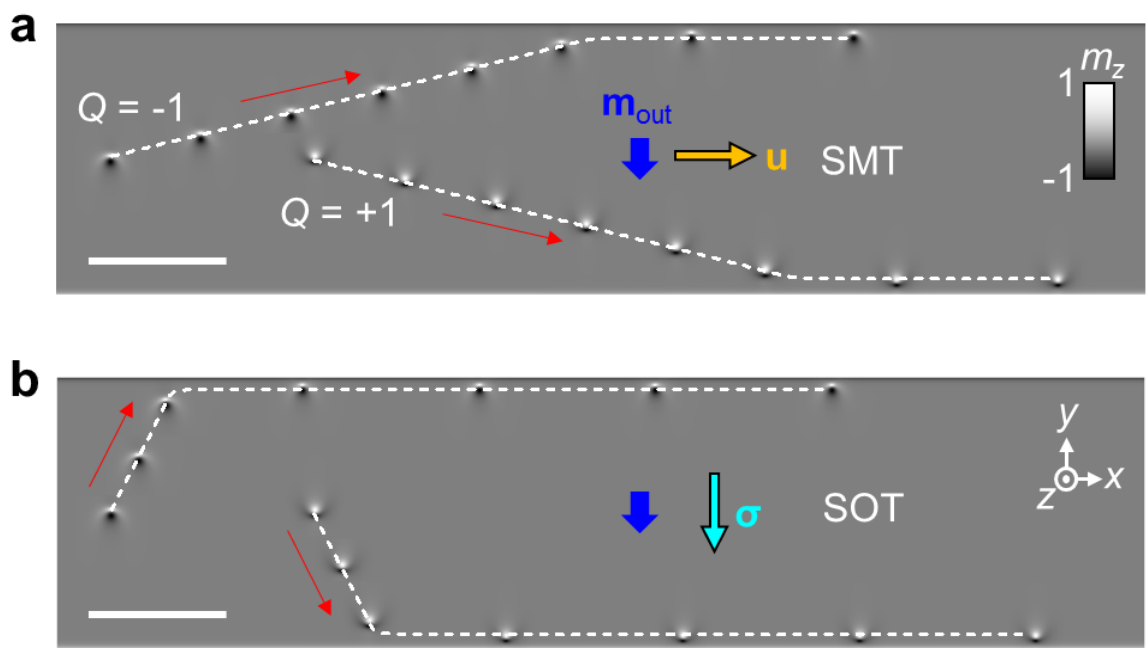

FIG. 4